\documentclass[11pt]{article}
\usepackage{times,cite}
\usepackage[T1]{fontenc}
\usepackage{amsmath,amssymb}
\usepackage[utf8]{inputenc}
\usepackage[colorlinks=true,linkcolor=black,citecolor=black,urlcolor=blue]{hyperref}
\usepackage{booktabs}
\usepackage{subcaption}
\usepackage{xcolor}
\usepackage[margin=2.5cm]{geometry}
\usepackage{graphicx}
\begin{document}

\title{MathModDB: A Database for Mathematical Models}
\date{}

\author{
Jochen Fiedler$^{1}$\thanks{Corresponding author: jochen.fiedler@itwm.fraunhofer.de, 
     phone +49\,631\,31600\,4771}
\and Christine Biedinger$^{1}$
\and Marco Reidelbach$^{2}$
\and Bj\"orn Schembera$^{3}$
\and Burkhard Schmidt$^{4}$
\and Aurela Shehu$^{4}$
\and Thomas Koprucki$^{4}$
}

\maketitle 
\vspace{-3em}
\begin{center}
{\small
$^{1}$Fraunhofer Institute for Industrial Mathematics, Kaiserslautern, Germany\\
$^{2}$Zuse Institute Berlin, Germany\\
$^{3}$Institute of Applied Analysis and Numerical Simulation, University of Stuttgart, Germany\\
$^{4}$Weierstrass Institute for Applied Analysis and Stochastics, Berlin, Germany
}
\end{center}
\vspace{1em}
\begin{abstract}
When researchers need a mathematical model for a research problem, they face a fragmented landscape: relevant formulas, quantities, assumptions, and model variants are scattered across publications and domain-specific conventions.
The Mathematical Models Database (MathModDB) addresses this challenge by providing a curated knowledge graph for mathematical models, deployed on the MaRDI Portal as part of the German National Research Data Infrastructure (NFDI).
Building on ontology designs presented in earlier work, this paper focuses on MathModDB as a publicly available service. It addresses researchers who use mathematical models in their work -- whether in applied mathematics, engineering, or the natural sciences.
We describe its deployment on the Wikibase-powered MaRDI Portal, report on its current scale, and demonstrate its practical use through a walkthrough of an electric discharge modeling use case from plasma physics.
We further discuss the ecosystem around MathModDB, including its connection to the MathAlgoDB knowledge graph for numerical algorithms and the MaRDMO documentation tool.
\end{abstract}
\smallskip
\noindent\textbf{Keywords:} research data management, mathematical models, knowledge graphs, FAIR principles, ontologies, scientific computing

\section{Introduction}
\label{sec:introduction}

Consider the following situation: a researcher wants to simulate a physical process -- say, an electric discharge in a plasma -- and needs to identify a suitable mathematical model.
Which formulas govern the process?
What quantities appear, and which are inputs, outputs, or parameters?
Are there simpler or more detailed model variants, and under what assumptions do they apply?
Which algorithms can solve the resulting computational task, and is there validated software available?

Answering these questions today typically requires extensive literature review, consulting with domain experts, and reconciling inconsistent or incomplete documentation.
Moreover, the same mathematical structures -- coupled PDE systems comprising conservation laws, constitutive relations, transport equations, couplings to external fields, etc. -- may appear across several scientific domains, yet this shared foundation is rarely made explicit in a way that allows discovery across disciplinary boundaries.
Even within a single academic discipline, model variants may differ subtly in their formulations or assumptions while being referred to by similar names, making comparison and reproduction of simulation results unnecessarily difficult.

Existing approaches to representing mathematical knowledge semantically -- such as OMDoc~\cite{Kohlhase2006} for mathematical documents, the OntoMath family of ontologies~\cite{Elizarov2017} for educational settings, or Wikidata~\cite{Vrandevic2014} for general-purpose knowledge representation -- focus on mathematical concepts or document structures rather than on models as research data.
Domain-specific ontologies for mathematical models exist in areas such as plasma physics~\cite{Snytnikov2020} or biology~\cite{Inizan2021}, but these are tailored to individual fields and do not provide a general, cross-disciplinary framework.

Within the Mathematical Research Data Initiative (MaRDI)~\cite{MaRDI2022}, a consortium of the German National Research Data Infrastructure (NFDI)~\cite{NFDI}, mathematical models are considered part of the broader landscape of mathematical research data~\cite{Koprucki2016,Koprucki2016b}. MaRDI aims to make mathematical research data \emph{FAIR (Findable, Accessible, Interoperable, Reusable)}~\cite{Wilkinson2016}. In this context, we have developed the Mathematical Models Database (MathModDB) to address the cross-disciplinary knowledge representation of mathematical models.
MathModDB is a curated knowledge graph that provides structured, semantic descriptions of mathematical models -- their formulas, quantities, assumptions, computational tasks, and links to literature.
The ontology design and its connection to other databases like the Mathematical Algorithms Database (MathAlgoDB), the knowledge graph for numerical algorithms, have been described in earlier work~\cite{Schembera2023_CoRDI, Schembera2024_MTSR, Schembera2025_MTSR, Schembera2025_CoRDI}.

This contribution focuses on MathModDB as a \emph{practical tool} that is publicly available to researchers at all levels -- from students to experts -- across applied mathematics, engineering, and the natural sciences.
Since 2025, the knowledge graph is deployed on the MaRDI Portal~\cite{MaRDIPortal,Schubotz2023}, a Wikibase-powered platform that serves as the central access point for all MaRDI services and data, including publications, software, workflows, and -- through MathModDB -- mathematical models.
We present the current state of the database -- covering more than 200 models with over 24\,000 semantic statements -- and demonstrate its use through a detailed walkthrough of an electric discharge modeling use case from plasma physics in Sec.~\ref{sec:portal}.
We further describe the ecosystem around MathModDB in Sec.~\ref{sec:ecosystem}, including its connections to MathAlgoDB, MaRDMO~\cite{Reidelbach_2025,Reidelbach2026a,Reidelbach2026b} -- a plugin for the Research Data Management Organiser (RDMO) that enables structured documentation of mathematical research data -- and domain-specific knowledge graphs.

\section{The MathModDB Ontology}
\label{sec:ontology}

MathModDB defines a data model for the semantic representation of mathematical models.
The ontology was originally developed in OWL/RDF~\cite{MathModDB_Zenodo} and has since been exported to the MaRDI Portal, where it is maintained and extended as a Wikibase-based knowledge graph (cf. Section~\ref{sec:portal}).
Detailed accounts of the ontology design and its evolution can be found in~\cite{Schembera2024_MTSR, Schembera2025_MTSR}.

\begin{figure}[t]
\centering
\includegraphics[width=\textwidth]{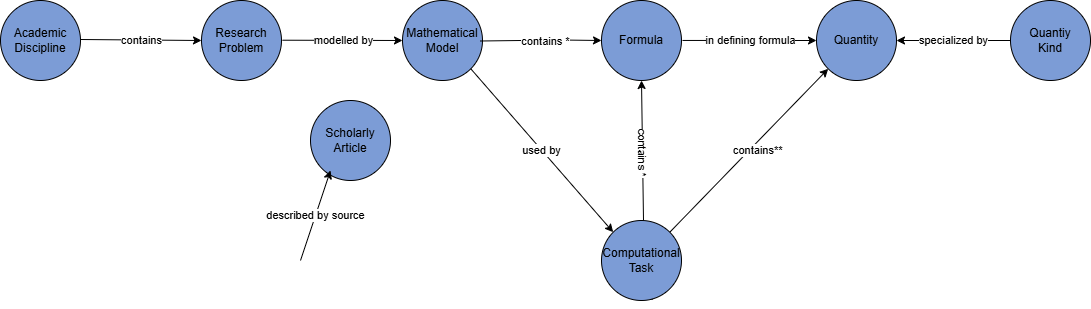}
\caption{Schematic overview of the MathModDB data model.
}
\label{fig:ontology}
\end{figure}

The central idea, inspired by Model Pathway Diagrams~\cite{Kohlhase2017, Koprucki2018}, is to decompose mathematical models into their constituent parts and the semantic relations between them.
The data model (cf. Figure~\ref{fig:ontology}) captures a chain from the application context to the mathematical details: an \emph{Academic Discipline} contains a \emph{Research Problem}, which is modeled by a \emph{Mathematical Model}.
Each model contains \emph{Formulas} (e.\,g., equations given as \LaTeX{} formulas), which in turn reference \emph{Quantities} and \emph{Quantity Kinds} (linked to the Quantities, Units, Dimensions and Data Types (QUDT) vocabulary~\cite{QUDT}).
\emph{Computational Tasks} specify concrete problems derived from a model -- for instance, determining particle densities for given boundary conditions -- by declaring which quantities serve as inputs, outputs, constants, or parameters.
Finally, \emph{Scholarly Articles} provide bibliographic provenance.

Two features of the data model deserve particular emphasis, as they shape how MathModDB is used in practice.
First, models are organized in \emph{specialization hierarchies}: a general model (e.\,g., Newton's second law) can be specialized into variants (e.\,g., free fall with or without air drag) by adding assumptions.
On the MaRDI Portal, these hierarchies are represented using Wikibase qualifiers that annotate the specialization relation with the relevant assumptions.
Second, the \emph{Computational Task} class serves as the interface to the algorithmic world: each task in MathModDB can be linked to a corresponding \emph{Algorithmic Task} in the MathAlgoDB knowledge graph~\cite{MathAlgoDB_Zenodo}, connecting model knowledge to available solution algorithms.

\section{MathModDB on the MaRDI Portal}
\label{sec:portal}

\subsection{Deployment and Portal Representation}
\label{subsec:deployment}

MathModDB is deployed on the Wikibase-powered MaRDI Portal\footnote{\url{https://portal.mardi4nfdi.de/wiki/Portal}}~\cite{Schubotz2023}, where each entity receives a persistent MaRDI QID enabling unambiguous identification and citation.
The portal provides several access modes: a web-based browsing interface with dedicated community pages for MathModDB, a SPARQL endpoint for structured queries against the full MaRDI Knowledge Graph, and the standard Wikibase API for programmatic access.
Additionally, links to external resources such as zbMATH Open, swMATH, the Digital Library of Mathematical Functions, QUDT and Wikidata are maintained, complying with the Linked Open Data (LOD) principles.

The portal user interface does not mirror the ontology one-to-one.
Instead, semantic relations are exposed through different UI components such as metadata badges, contained-entities tables, identifier sections, and further-links boxes.
Table~\ref{tab:ui_mapping} provides a systematic mapping, and we illustrate the portal representation through a use case walkthrough in Section~\ref{subsec:usecase}.

\begin{table}[t]
\centering
\caption{Mapping of semantic relations to portal UI sections, illustrated for the Fluid Poisson model.}
\label{tab:ui_mapping}
\begin{tabular}{p{3.8cm}p{3.2cm}p{5.5cm}}
\toprule
\textbf{Semantic relation} & \textbf{Portal UI section} & \textbf{Example} \\
\midrule
model addresses research problem & further links & fluid Poisson model $\to$ electric discharge \\
model contains formulas & contained entities & continuity eq., Poisson's eq., particle flux, \ldots{} (7 total) \\
formula contains quantities & symbol annotations in formula rows & $\phi$ $\to$ electric potential, $\rho$ $\to$ charge density \\
model has assumptions & assumptions & classical approx., continuum approx., local approx. \\
model is characterized by tags & top metadata badges & deterministic, dynamic, nonlinear, \ldots \\
model is linked to tasks & computational tasks & determine particle densities \\
model links to software & software & FEniCS \\
quantity has external identifiers & available identifiers & permittivity of vacuum $\to$ Wikidata, QUDT \\
\bottomrule
\end{tabular}
\end{table}

\subsection{Current Scale and Entry Points}
\label{subsec:scale}
 
Table~\ref{tab:statistics} summarizes the current content of MathModDB.
The content covers a wide range of application domains, including but not limited to computational mechanics (in cooperation with NFDI4Ing\footnote{The NFDI consortium for \emph{research data management in engineering sciences}, \url{https://nfdi4ing.de/}} and the Bundesanstalt f\"ur Materialforschung, Berlin), plasma physics (in cooperation with the Leibniz Institute for Plasma Science and Technology, Greifswald), and semiconductor device simulation.

For new users of MathModDB, the portal provides curated entry points in the form of \emph{featured research problems}\footnote{\url{https://portal.mardi4nfdi.de/wiki/MathModDB\#Featured_research_problems}}. 
These are research problems that have been documented in particular depth, with complete model hierarchies, formulas, quantities, and links to literature or software.
Currently featured research problems include, among others, charge transport in semiconductor devices and electric discharge in non-thermal plasmas.
We use the latter as a running example to illustrate how MathModDB represents model knowledge in practice.
 
\begin{table}[t]
\centering
\caption{Current content of the MathModDB knowledge graph.}
\label{tab:statistics}
\begin{tabular}{lr}
\toprule
\textbf{Class} & \textbf{Count} \\
\midrule
Academic Disciplines & 50 \\
Research Problems & 107 \\
Mathematical Models & 227 \\
Computational Tasks & 173 \\
Formulas & 708 \\
Quantities & 792 \\
Quantity Kinds & 68 \\
\midrule
\textbf{Total number of individuals} & \textbf{2\,125} \\
\textbf{Total number of statements} & \textbf{24\,762} \\
\bottomrule
\end{tabular}
\end{table}

\subsection{Use Case: Electric Discharge Modeling}
\label{subsec:usecase}

The electric discharge use case was developed in cooperation with the Leibniz Institute for Plasma Science and Technology (INP Greifswald) and the NFDI4BIOIMAGE consortium.
It demonstrates how MathModDB represents a non-trivial model from plasma physics, covering the full chain from application context to computational tasks.

\begin{figure}[p]
\centering

\begin{subfigure}[t]{\textwidth}
\includegraphics[width=0.54\textwidth]{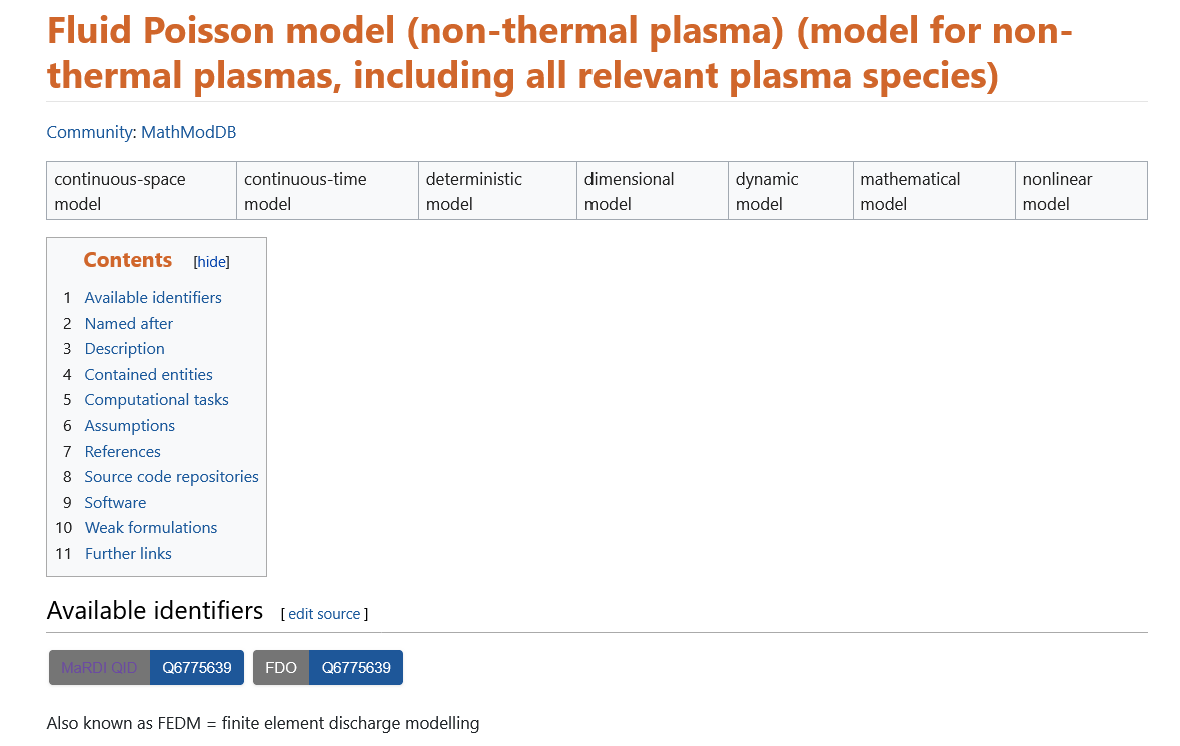}
\caption{Header section with model name, classification badges, and identifiers.}
\end{subfigure}

\bigskip

\begin{subfigure}[t]{\textwidth}
\includegraphics[width=0.54\textwidth]{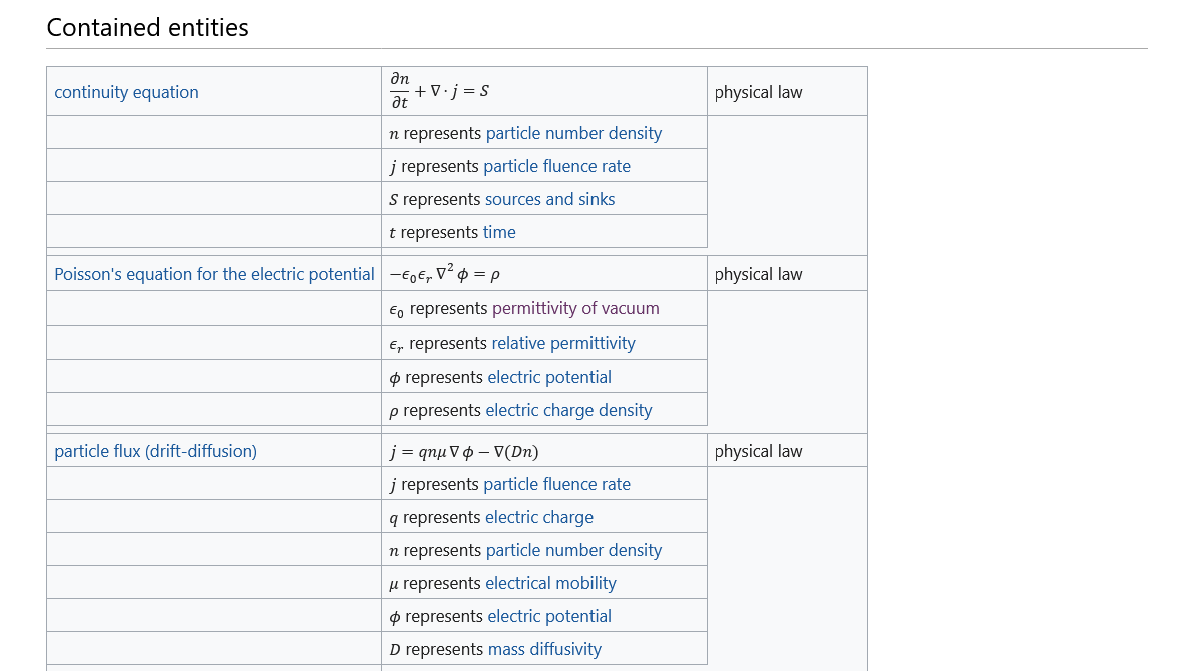}
\caption{Contained entities: formulas with \LaTeX{} rendering, type classification, and symbol-to-quantity mappings via qualifiers.}
\end{subfigure}

\bigskip

\begin{subfigure}[t]{\textwidth}
\includegraphics[width=0.54\textwidth]{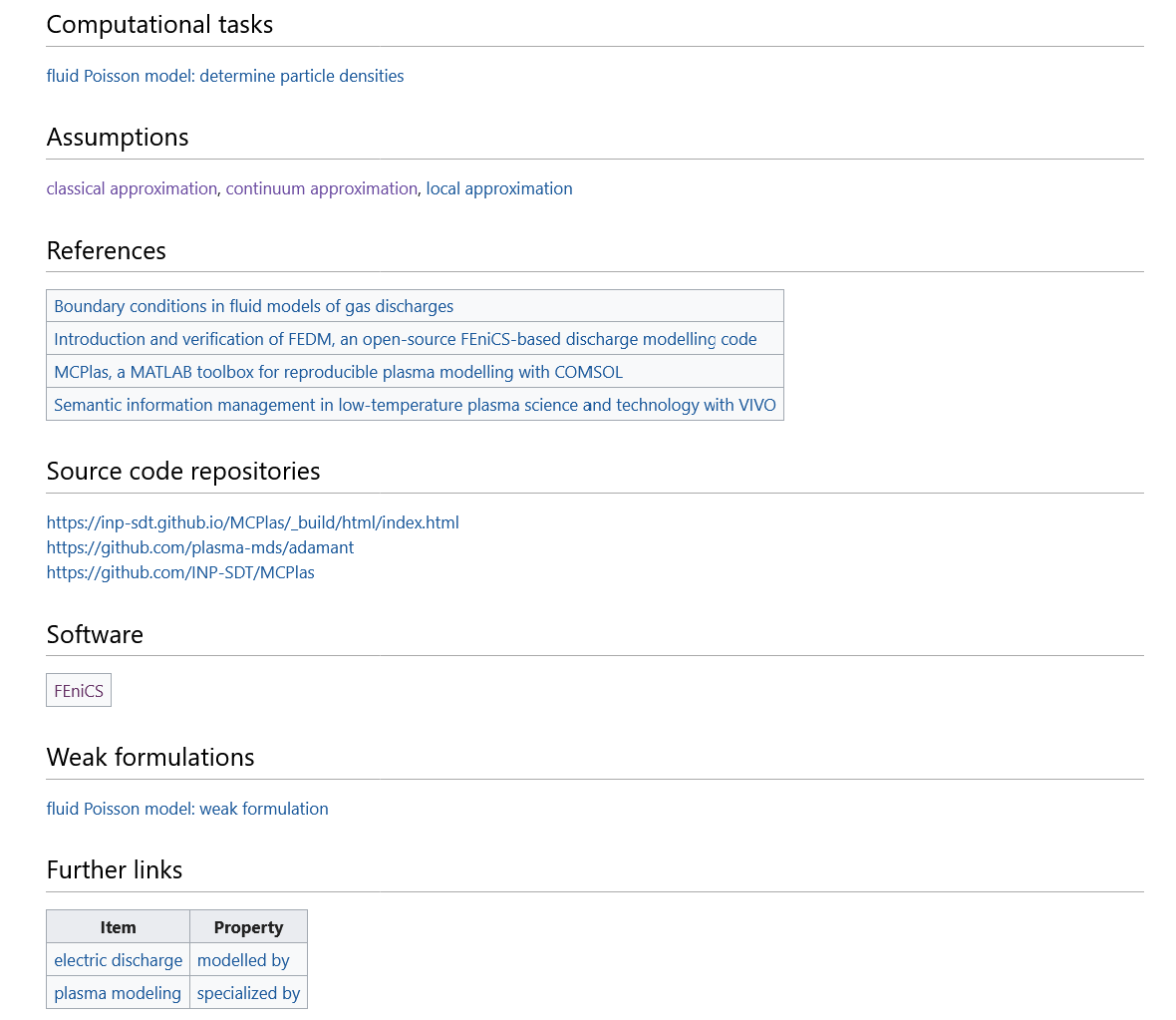}
\caption{Computational tasks, assumptions, references, software links, and further links (connecting the model to its research problem and academic discipline).}
\end{subfigure}

\caption{Portal representation of the Fluid Poisson model (non-thermal plasma) in MathModDB, shown as three annotated sections of the model page. Together, these sections expose the semantic relations listed in Table~\ref{tab:ui_mapping}.}
\label{fig:portal_model}
\end{figure}

\paragraph{The Model and its Portal Representation.}
The central entity in this use case is the "Fluid Poisson model in non-thermal plasma"\cite{Jovanovic2023}, a system of balance equations for charged and neutral species along with drift-diffusion equations describing their transport coupled with Poisson's equation for the electric potential.
Figure~\ref{fig:portal_model} shows the model's portal page, subdivided into three sections to illustrate how MathModDB exposes model metadata through the portal interface.\footnote{The model has the MaRDI QID Q6775639 \url{https://portal.mardi4nfdi.de/wiki/Item:Q6775639}.}
Table~\ref{tab:ui_mapping} provides a systematic mapping between semantic relations in the ontology and the UI sections where they appear.

The model properties (spatial/temporal continuity, determinism, linearity, etc.) provide a structured basis for comparison with other models.
The assumptions -- classical approximation, continuum approximation, and local approximation -- are listed explicitly, making the model's scope of validity transparent and comparable with alternative formulations.
The page also links to software implementing the model, source code repositories, and a weak formulation variant of the underlying PDEs which are the basis for finite element implementations.

\paragraph{Model Hierarchies.}
MathModDB organizes models in specialization hierarchies that reflect the physical assumptions under which one model reduces to another.
In the electric discharge context, the Fluid Poisson model is itself part of a broader family: it is contained in the more general class of plasma modeling approaches, while the reaction kinetics model appears as one of its contained sub-models.
On the portal, these hierarchical relations are visible in the "further links" section, where parent models, specializations, and contained sub-models are listed.
This structure allows researchers to navigate upward toward more general formulations or downward toward specific variants tailored to particular applications, with the qualifying assumptions documented at each step.

\paragraph{Formulas and Quantities.}
The model contains seven formulas, each documented with its \LaTeX{} formula, its type, and the symbols appearing in it.
On the model page, the formulas are listed in the table of contained entities (cf.\ Figure~\ref{fig:portal_model}b).
As representative examples, we highlight three physical laws:
\begin{itemize}
    \item The \emph{continuity equation}: $$\frac{\partial n}{\partial t} + \nabla \cdot j = S;$$
    \item \emph{Poisson's equation for the electric potential}: $$-\varepsilon_0\varepsilon_r \nabla^2 \phi = \rho;$$
    \item the \emph{particle flux}: $$j = qn\mu\nabla \phi - \nabla(Dn).$$ 
\end{itemize}
In these laws, $j$ represents the particle fluence rate, $n$ the particle number density, $S$ sources and sinks, $t$ the time, $\varepsilon_0$ the permittivity of vacuum, $\varepsilon_r$ the relative permittivity, $\phi$ the electric potential, $\rho$ the electric charge density, $q$ the electric charge, $\mu$ the electrical mobility and $D$ the mass diffusivity.
The remaining formulas include a Robin boundary condition for the particle flux at electrodes, the electron energy flux and its continuity equation, and a reaction kinetics model for the involved chemical species.
Each formula is classified by type (e.\,g., physical law, boundary condition), and each symbol is linked via qualifiers to a Quantity.

The quantities themselves are documented as individual entities with their own portal pages.
For instance, the permittivity of vacuum has its ISQ dimension, links to Wikidata and a QUDT constant ID, ensuring unambiguous identification across models and disciplines.
This level of detail means that quantities are not merely textual labels, but semantically grounded LOD entities that can be reused whenever the same physical quantity appears in a different model.

\paragraph{Application Context.}
From the model, the knowledge graph connects to the research problem "electric discharge", which in turn is contained in the academic discipline of plasma physics, specifically "nonthermal plasma".
The research problem entity includes a textual description, references to relevant publications, and lists all mathematical models associated with it.
This makes it possible to discover alternative or complementary modeling approaches for the same phenomenon -- for instance, different levels of complexity in plasma modeling, such as kinetic vs.\ fluid descriptions.

\paragraph{Computational Tasks.}
The model is associated with the computational task "determine particle densities", classified as a linear task. After changing to a weak formulation and choosing a proper discretization, the task of solving the above equations reduces to a linear system $Ax = b$, corresponding to the algorithmic task "linear system of equations (Ax=b)" in MathAlgoDB, giving a choice of algorithms for its solution. Note that establishing such links between MathModDB and MathAlgoDB is ongoing work.

\paragraph{Summary.}
This walkthrough demonstrates how MathModDB condenses model knowledge into an explorable, interlinked structure.
A researcher encountering the term "fluid Poisson model" can find the governing formulas, understand which quantities are involved, trace the assumptions and specializations, identify relevant literature and software, and -- through the link to MathAlgoDB -- discover suitable numerical algorithms.

\section{Ecosystem Connections}
\label{sec:ecosystem}

MathModDB is embedded in a broader ecosystem of tools for mathematical research data management, making it a central part of mathematical knowledge management.

\paragraph{MathAlgoDB}
MathAlgoDB~\cite{MathAlgoDB_Zenodo} is a complementary knowledge graph for numerical algorithms, also developed within MaRDI.
The connection between MathModDB and MathAlgoDB is established through the \emph{Computational Task} in MathModDB and the \emph{Algorithmic Task} in MathAlgoDB~\cite{Schembera2025_CoRDI}: the former describes \emph{what} needs to be computed, the latter \emph{how} it can be computed.
This interface allows users to navigate from a mathematical model to available solution algorithms, software implementations, and benchmarks, without either knowledge graph needing to duplicate the other's content.

\paragraph{MaRDMO.}
To lower the barrier for community contributions, the MaRDMO plugin~\cite{Reidelbach_2025,Reidelbach2026a,Reidelbach2026b} extends the Research Data Management Organiser (RDMO)~\cite{Engelhardt2017} with mathematics-specific questionnaires.
MaRDMO guides researchers through the documentation of mathematical models, algorithms, and interdisciplinary workflows in a structured format and ingests the results directly into the MaRDI Portal.
A detailed description of MaRDMO, including a Stokes-Darcy use case demonstrating the documentation and search of algorithms, mathematical models, and interdisciplinary workflows, is provided in the complementary paper~\cite{Reidelbach2026_PAMM} in these proceedings.

\paragraph{Interoperability and Outlook.}
At the data level, MathModDB entities are linked to Wikidata, QUDT, and publication databases. By using the Wikibase technology, it is possible to incorporate and link additional data sources easily. 
Ongoing work aims to connect MathModDB with domain-specific knowledge graphs such as PlasmaKG, to align with the NFDI Core Ontology~\cite{Bruns2024}, to integrate with workflow systems for provenance tracking in computational experiments and to incorporate MathModDB into the TS4NFDI terminology service~\cite{Zapilko2025}.

\section{Conclusion and Outlook}
\label{sec:conclusion}

We have presented MathModDB as a deployed service on the MaRDI Portal, focusing on its practical use for discovering, documenting, and reusing mathematical model knowledge.
With more than 2\,100 entities and 24\,000 semantic statements covering over 200 models from diverse application domains, MathModDB provides a practical resource that implements the FAIR principles for mathematical models.

Through its connection to MathAlgoDB and MaRDMO, MathModDB is embedded in a growing ecosystem that spans the modeling-simulation-optimization workflow from research problems to solution algorithms.
Several challenges remain: the current expert-curated ingestion process does not scale to all disciplines, and strategies for semi-automated extraction of model metadata from publications are being investigated.
The handling of discretization -- an essential step in connecting continuous models to discrete algorithms -- is currently only implicitly represented and requires conceptual refinement.

We invite the applied mathematics community to explore MathModDB at \url{https://portal.mardi4nfdi.de/wiki/MathModDB} and to contribute models, feedback, and use cases.

\paragraph{Acknowledgements.}
The work has been funded by the DFG (German Research Foundation), project number 460135501, NFDI 29/1 "MaRDI -- Mathematische Forschungsdateninitiative".
The authors thank Markus Becker (INP Greifswald) for the collaboration on the electric discharge use case.

\vspace{\baselineskip}
 \bibliographystyle{unsrt}
  \bibliography{references}

\end{document}